\documentclass[pra,superscriptaddress, nofootinbib, showpacs, twocolumn]{revtex4}
\usepackage{graphicx}       
\usepackage{bm}         
\usepackage{amsfonts}
\usepackage{amsmath}        
\usepackage{latexsym}
\usepackage{amssymb}

\begin{document}

\title{Information communicated by entangled photon pairs}
\author{Thomas Brougham and Stephen M. Barnett\\ 
Department of Physics, University of Strathclyde, Glasgow, G4 0NG, U.K.}

\begin{abstract}
A key goal of quantum communication is to determine the maximum number of bits shared between two quantum systems.  
An important example of this is in entanglement based quantum key distribution (QKD) schemes.  A realistic treatment of this general communication problem must take account of the nonideal nature of the entanglement source and the detectors.  The aim of this paper is to give such a treatment.  
We obtain analytic expression for the mutual information in terms of experimental parameters.  The results are applied to communication schemes that rely on spontaneous parametric down conversion to generate entangled photons.  We show that our results can be applied to tasks such as calculating the optimal rate of bits per photon in high dimensional time bin encoded QKD protocols (prior to privacy amplification).  A key finding for such protocols is that by using realistic experimental parameters, one can obtain over 10 bits per photon.  We also show how our results can be applied to characterize the capacity of a fibre array and to quantify entanglement using mutual information.
\end{abstract}
\maketitle

\section{Introduction}
A central concern of quantum information theory is to determine the maximum amount of information that can be shared between two quantum systems.  The shared information gives an indication of the quantum correlations and is thus of fundamental interest \cite{BDSW,BBPS,indexcorr,hallcorr}.  In addition to this, knowledge of the shared information determines the performance of quantum communications \cite{bennettshor} and is vital for many proposed applications of quantum information \cite{barnett,nch}.  One important example of this is in quantum key distribution which, in one of the seminal insights within the field of quantum information,  was the realisation that quantum mechanics allows for the secure distribution of cryptographic keys \cite{bennettbrassard,Bennett92,PBC2000,qkdreview,NRA}.  An important class of quantum key distribution (QKD) scheme are those that use entanglement \cite{ekert,BBM92,etime,largealpha}.  The security of entanglement based QKD is based on the nonlocality of quantum mechanics.

An important issue to address for QKD is determining how many bits of secret key are distributed.  A prerequisite for this calculation is determining the maximum number of unsecured bits that the two parties can share.  This is the information contained in the raw keys.  To fully extract this information, error correction is necessary, followed by   privacy amplification to minimise the information an eavesdropper might have access to \cite{privamp}.  If one is looking to optimise both the error correction and privacy amplification protocols, then it is essential to know the mutual information shared between the raw keys.  This quantity will depend on both the detectors and the details of the source of the entangled states.  Realistic physical models of each of these components is thus vital.


The effects of losses, inefficiencies and imperfect sources have been studied previously \cite{loss1}.  In particular, spontaneous parametric down conversion sources with losses have been studied with regards to QKD \cite{loss2,loss3}.  The existing work, however, has concentrated on the case where the information is encoded in the polarisation degree of freedom.  As a result, this work cannot be applied to communications protocols that use high dimensional entangled states with the aim of encoding multiple bits on each photon. 

In this paper we determine the maximum shared information for realistic quantum sources and detectors.  As most experimental implementations of quantum communication protocols are optically based, the examples we study will be limited to optical systems.  The natural question to ask is: what is the maximum shared information per photon?  The main result of our paper will be to determine this for general but experimentally relevant conditions.  This result is broader in its scope than earlier findings, such as \cite{loss2,loss3}. In particular, our results apply not only to information encoded in polarization, but also to information encoded in high dimensional entangled degrees of freedom.  This fact is illustrated by applying our findings to different experimental setups, which include, but are not limited to, QKD experiments.  Our results can thus be used to optimise a given experimental procedure.  In particular, one can determine the power at which to operate a pump laser driving a spontaneous parametric down conversion source, so as to maximise the mutual information.

The paper is organized in the following way.  In section 2 we discuss the general communication system that we consider.  Special attention is paid to time bin encoded communication protocols \cite{tb1,tb2,tb3,tb4}.  In sections 3 and 4  we construct simple mathematical models of the source, channels and imperfect photon detectors.  The mutual information is calculated in section 5 and 6 for sources that produce entangled pairs with a Poissonian distribution.  In section 7 we discuss the relevance of our results to other tasks such as characterizing the capacity of a fibre array and information theoretic approaches to quantifying entanglement.  Finally, our results are discussed in section 8.  

\section{Communicating shared bits using entanglement}

The fundamental communications problem is distributing a shared string of bits between two parties, called Alice and Bob.  One way of using quantum mechanics to achieve this is to use an entangled state, i.e. one of the form
\begin{equation}
|\Psi\rangle_{AB}=\sum_k{c_k|\varphi_k\rangle_A|\varphi'_k\rangle_B},
\end{equation}
where $\langle\varphi_i|\varphi_j\rangle=\langle\varphi'_i|\varphi'_j\rangle=\delta_{ij}$.  It can also be useful to consider hyper-entangled states that have entanglement between multiple degrees of freedom \cite{hyper1,hyper2}.  A string of bits can then be generated from the correlation between one of the entangled degrees of freedom.  If one is looking to communicate multiple bits per entangled state, then this requires high dimensional entanglement.  The idea is best illustrated by some simple examples.

Using spontaneous parametric down conversion (SPDC) one can produce pairs of photons that are entangled both in their polarization and in time / frequency \cite{etime,spaciotemp}.  It is convenient to express such states using continuous time creation and annihilation operators, which satisfy the commutator relation $[\hat a(t),\hat a^{\dagger}(t')]=\delta(t-t')$ \cite{loudon}.  The down converted state thus has the form
\begin{eqnarray}
|\Psi\rangle_{AB}=\frac{1}{\sqrt{2}}\left(|HH\rangle_{AB}+|VV\rangle_{AB}\right)\nonumber\\
\otimes\int{dt_1dt_2g(t_1,t_2)\hat a^{\dagger}_A(t_1)\hat a^{\dagger}_B(t_2)|0\rangle},
\end{eqnarray}
where $H$ and $V$ respectively denote horizontal and vertical polarization, $\hat a^{\dagger}_{A,B}(t)$ is the creation operator for Alice / Bob's photons and $g(t_1,t_2)$ is a normalized function that is zero when $|t_1-t_2|$ become sufficiently large.  This means that if Alice detects a photon at time $T$, then Bob should also detect his photon at a time close to $T$.  This correlation can be used to form a random bit string shared between Alice and Bob.  One way of doing this is to divide the photon's possible arrival times into discrete time bins and record whether photons are found in each time bin.  The measurement record of $M$ time bins can then be used, at least in principle, to construct a string of $M$ bits.  The zeros and ones of the bit string would respectively correspond to the absence or detection of photons within a time bin. 

A second example is to use the spatial modes of entangled photons generated by SPDC.  It has been shown, for example, that the down converted photons are entangled in their orbital angular momentum and angular position \cite{Mair,gotte,leach}.  The fact that angular momentum requires an unbounded Hilbert space means that measuring the angular momentum of the photons enables multiple bits to be extracted per photon pair.  For example, if an experiment can distinguish between $M$ different angular momentum eigenstates, then one can extract up to $\log_2(M)$ bits per photon pair.

It is worthwhile noting the connection between this communication problem and QKD.  For a given quantum key distribution protocol, one wishes to distribute a {\it secret} shared string of bits.  This requires us to outline a mechanism for checking the security of the bits.  A common approach is to measure in two mutually unbiased bases and then publish a random sample of Alice and Bob's measurement results.  The presence of an eavesdropper can then be established by looking at the cases where Alice and Bob measured in the same basis.  In the absence of an eavesdropper, these measurement results should be correlated.  The use of incompatible bases and implementing security checks necessarily reduces the communication rate.

The communication problem we are studying corresponds to determining the shared information when Alice and Bob both measure in the same basis.  It is important to note that this is not the number of shared secret bits.  To determine this one would need to find out the number of bits lost due to the additional constraints of security.  For details on how the security of QKD decreases the shared information see \cite{nch,privamp}.

\section{Modelling the photon source, detectors and losses}

The first part of the system that we will look at is the source of the entangled photons.  We first assume that the source generates entangled pairs of photons.  If four photons are generated, then this will correspond to two pairs of entangled photons\footnote{In practice this condition may have to be relaxed slightly.  Alternatively, one might filter the output of the source so as to post-select only those situations when pairs were emitted.}.  The next assumption is we have a probability $P(m)$ of producing $m$ photon pairs at a time and that this is independent of the number of photon pairs produced earlier.  For the example of time bin encoded photons, $P(m)$ corresponds to  the probability of producing $m$ photon pairs within a given time window.   When the photons are generated by SPDC, $P(m)$ can be approximated by a Poissonian distribution \cite{poissonref}.  
  
The next aspect that we consider is the measurement process.  We consider only measurements that are realised either by the detection of photons within time bins or within discrete spatial locations or modes.  This does not limit us, however, to considering only temporal or spatial encoded information.  The reason for this is that measurements of other degrees of freedom can be converted into measurements of either temporal or spatial degrees of freedom.  A common example of this is measuring polarization by using a polarizing beam splitter. This converts information about the polarization into information about the spatial location of a photon.  Similarly, one can convert measurements of different quantities into measurements of the time of arrival of a photon.  For example, the spatial position of photons in an optical field have been measured using a time multiplexing fibre array.  Different position at which the photon could be found where converted into different time windows in which the photon could be detected \cite{PadgettBuller}.  The advantage of this sort of experimental procedure is that only one photon detector is required to detect many different spatial modes.  

Another assumption of our analysis is that one is equally likely to obtain any of the measurement outcomes.  For example, in a time-binned system this would correspond to it being equally likely that photons are sent in each time bin.  While the assumption is reasonable in the context of time-binned communication systems, it can appear prohibitive in other situations.  The reason for making this assumption is that the aim of our analysis is to determine the effect of inefficiencies in the source, channel and detectors.  For this reason we consider only the simplest possible form for the entangled photonic states.  More complex systems can, however, still be approximated by our analysis.  For example, the effective Schmidt modes of an SPDC state with large Schmidt number can, to a good approximation, be taken to be equiprobable \cite{schmidtpaper}. 

As our aim is to encode information on each photon pair and not in the number of photons, we shall consider only threshold detectors that do not resolve photon number.  Consider first the situation where we have no losses.  This would correspond to no transmission losses and ideal detectors that detect all photons incident on them.  The probability for Alice or Bob's detector to fire is 
\begin{equation}
\pi(c)=\sum_{m=1}^{\infty}{P(m)},
\end{equation}
where $\pi(c)$ is the probability for the ideal detector to click and $P(m)$ is again the probability to find $m$ photons within each time bin or spacial mode.  In a real experiment, however, detectors do have losses.  One can model an inefficient detector as an ideal detector with a lossy medium in front of it \cite{john,detect}.  For our purposes, the lossy medium can be viewed as a beam splitter with transmission coefficient $\sqrt{\eta_d}$.  The losses will thus correspond to photons being reflected at the beam splitter.  The square of the transmission coefficient, $\eta_d$, corresponds to the efficiency of the detector.  It is clear that a single photon incident on the nonideal detector will be detected with probability $\eta_d$.  

In addition to the photons lost by the detectors there will be losses during transmission.  This can again be modelled by a beam splitter, where the transmission coefficient is $\sqrt{\eta_l}$.  It is convenient to incorporate both of these sources of losses into a single efficiency for the system.  This would correspond to a beam splitter with transmission coefficient $\sqrt{\eta}$, where $\eta=\eta_d\eta_l$.  The total efficiency $\eta$ again gives the probability for any given emitted photon to be detected.


One further source of loss is in cross talk between multiple modes.  This would be an issue if one is transmitting the information in spatial modes.  The effect of cross talk is to cause photons to be lost from one mode and appear in a different one.  A simple way to account for this in our current model is to adjust the efficiency $\eta$ to include loss from cross talk.  The effect of the photons appearing in a different mode can then be dealt with by increasing the effective dark count rate.

\section{Joint detection probability} 
To calculate the mutual information we need to determine the joint probability for Alice and Bob to obtain the same measurement outcome.  The key mathematical tool in our analysis is the moment generating function \cite{barnettradmore}.  If $P(n)$ is the probability that a pulse contains $n$ photons, then we define the moment generating function to be
\begin{equation}
\label{onemoment}
M(\mu)=\sum_{n=0}^{\infty}{P(n)(1-\mu)^n}.
\end{equation}
Moment generating functions have many useful properties, the simplest of which is
\begin{equation}
\label{probnphotons}
P(n)=\frac{1}{n!}\left(-\frac{d}{d\mu}\right)^nM(\mu)\Big|_{\mu=1}\,.
\end{equation}
It is straightforward to generalize the definition of $M(\mu)$ to a pair of pulses:
\begin{equation}
\label{twomoment1}
M(\mu,\xi) = \sum_{m,n=0}^\infty P(m,n) (1 - \mu)^m(1 -\xi)^n \, .
\end{equation}
In our case the pair of pulses correspond to the entangled signal and idler beams for a single time bin or spatial mode.  The number of photons in each pulse is the same, hence 
\begin{equation}
P(m,n)=P(n)\delta_{m,n}.  
\end{equation}
The merit of using the moment generating functions is that it is easy to account for losses \cite{barnettradmore}. 

If we suppose that Alice and Bob both have identical detectors and that both their channels have the same losses, then we can assign to both parties the same total efficiency $\eta$. The generating function is
\begin{equation}
\label{twomoment2}
M_{loss}(\mu,\xi) = \sum_{m=0}^\infty P(m) (1 - \eta\mu)^m(1 -\eta\xi)^m.
\end{equation}
Consider a particular measurement outcome.  This will either correspond to a spatial location at which photons can be detected or a particular time bin in which the photons can be found.  Let $c$ and $0$ denote the detectors registering a click and not registering a click, respectively.  The joint probability for Alice and Bob to both get the same measurement outcome is $\mathcal{\pi}^{AB}(i,j)$, where $i,j\in\{0,c\}$.  In the ideal case $\mathcal{\pi}(0,c)=\mathcal{\pi}(c,0)=0$, i.e. Alice and Bob would either both detect photons or neither would detect any.  This is not the case, however, when losses are present.  In the absence of dark counts the probabilities are 
\begin{eqnarray}
\label{piprobs}
\pi^{AB}(0,0) &=& M_{loss}(1, 1),\nonumber \\
\pi^{AB}(c,0)&=&\sum_{l=1}^\infty \frac{1}{l!}\left. \left(-\frac{d}{d \xi}\right)^l M_{loss}(1, \xi)\right|_{\xi = 1},\nonumber\\
\pi^{AB}(0,c)&=&\pi^{AB}(c,0),\nonumber\\
\pi^{AB}(c,c)&=&\sum_{n=1}^\infty P(n)\left[1 - (1 - \eta)^n\right]^2.
\end{eqnarray}

The effect of the detectors registering dark counts can easily be modelled.  Let $q$ be the probability that, within a given period of time, Alice or Bob's detector fires when no photons are incident on it.  The joint probability, $\mathcal{P}^{AB}(i,j)$, will thus be 
\begin{eqnarray}
\label{fullprobs}
\mathcal{P}^{AB}(0, 0) &=& (1 - q)^2\pi^{AB}(0, 0) \nonumber \\
\mathcal{P}^{AB}(0, c) &=& (1 - q)\pi^{AB}(0, c) + (1-q)q\pi^{AB}(0, 0) \nonumber \\
&=& \mathcal{P}^{AB}(c, 0) \nonumber \\
\mathcal{P}^{AB}(c,c) &=& \pi^{AB}(c,c) + 2q\pi^{AB}(0,c) + q^2\pi^{AB}(0,0) \, .\nonumber\\
\end{eqnarray}
It is straightforward to verify that these probabilities sum to one.  The marginal probabilities for Alice or Bob's detector to fire are $\mathcal{P}(0) = (1-q)\left[\pi(0,0) + \pi(0,c)\right]$ and $\mathcal{P}(c)=1-\mathcal{P}(0)$.  We have thus obtain a general expression for the joint and marginal probability distributions.  These expressions are valid for any choice for the source probability $P(m)$ and consequently are not tied to one physical implementation.


\section{Mutual information}
In the classic paper of Shannon it was shown that the maximum amount of information that two parties can share is given by the mutual information \cite{shannon,EIT}.  For a joint probability distribution $\mathcal{P}^{AB}(i,j)$ with marginal probabilities $\mathcal{P}^A(i)$ and $\mathcal{P}^B(j)$, the mutual information is defined as
\begin{equation}
\label{hab}
H(A:B) = \sum_{i,j = 0,c}\mathcal{P}^{AB}(i,j)\log_2
\left(\frac{\mathcal{P}^{AB}(i,j)}{\mathcal{P}^A(i)\mathcal{P}^B(j)}\right)  \,. 
\end{equation}
The quantity given in equation (\ref{hab}) is the mutual information that Alice and Bob share when they obtain the same measurement outcome.  Consider a time bin encoded QKD protocol.  If $M$ time bins are used to create a key, then Alice and Bob will share a bit string of length $MH(A:B)$.  When one is interested in QKD, then number of shared bits will, of course, be reduced by the need to perform privacy amplification.

For communication in the quantum regime, it often important to know the number of shared bits per photon.  We must, however, be careful in how this quantity is defined.  One approach would be to divide the mutual information by the mean number of photon pairs produced.  If we denote the mean number of photon pairs by $\lambda$, then the information per photon is simply
\begin{equation}
\label{igen}
I_g(A:B)=\frac{H(A:B)}{\lambda}.
\end{equation}
The quantity $I_g(A:B)$ is the information per generated photon pair, however, not all generated photons are detected.  This fact suggests an alternative way to define the information per photon.  Instead of considering the information per generated photon pair, we can use the information per detected photon pair.  This is defined as 
\begin{equation}
\label{idet}
I_d(A:B)=\frac{H(A:B)}{\eta^2\lambda+q^2}.
\end{equation}


The formalism developed so far applies to any choice for $P(m)$.  We shall examine a concrete example, where $P(m)$ is a Poissonian distribution.   

\section{The mutual information for SPDC sources}
An important way of generating entangled photons is via 
the process of spontaneous parametric down conversion (SPDC).  In this approach a pump beam illuminates a nonlinear crystal.  Within the crystal, each pump photon can be converted into two lower frequency photons, referred to as signal and idler photons.  By careful choice of the systems parameters, one can arrange for the emitted photon pairs to be entangled.  This entanglement can be both in the polarisation and in the transverse spatial modes \cite{spdcentangle,spiral,filippo}.  This method of generating entangled photons has been used in many experimental realisations of QKD \cite{hyper1,qkdspdc}.

To illustrate how SPDC can be used to distribute a string of random bits consider the following two examples.  A time-binned protocol can be implemented using a pulsed laser that is shone at a nonlinear crystal.  The resulting down-converted photon pairs are entangled in time.  Distributing these photons to Alice and Bob allows them to construct a shared random string of bits.  The entangled time bin states needed for this protocol have already been experimentally generated \cite{tb1,tb2,tb3}.  The second example is to use the spatial modes of photons generated by SPDC.  By using a pump beam with an appropriate profile \cite{alison}, one can generate photon pairs that carry angular momentum in their transverse modes \cite{spiral,filippo,lesallen}.  The down converted photons will be entangled in the angular momentum degree of freedom.  A mode sorter can be used to separate some of the different angular momentum states into different spatial locations, where photon detectors are located.  In the limit of large Schmidt number, the effective modes of the photon pairs, will be approximately equiprobably \cite{schmidtpaper}.  The analysis we present will apply to both of these examples as well as many other situations where photons are generated by SPDC.

The probability distribution for $m$ pairs of photons to be generated by SPDC can be approximated by a Poissonian distribution 
\begin{equation}
\label{poisson}
P(m)=e^{-\lambda}\frac{\lambda^m}{m!}.
\end{equation}
It can easily be verified that $\lambda$ is the mean number of photon pairs.  
Using this distribution we can derive the mutual information for QKD schemes that use SPDC to generate pairs of entangled photons.  
The use of a Poissonian for the probability distribution is only valid when the initial laser pulses are not too short.  If we instead have short pulses, then it becomes necessary to use a thermal distribution, i.e. $P(m)=\lambda^m/(\lambda+1)^{m+1}$.  The moment generating function can again be calculated and the results of section IV can be used to calculated the new value for the mutual information \footnote{A quick calculation shows that for a thermal distribution, the moment generating function is $M_{loss}(\mu,\xi)=(1+\eta\lambda[\mu+\xi-\eta\mu\xi])^{-1}$.}. 

\begin{figure}
\center{\includegraphics[width=8.3cm,height=!]
{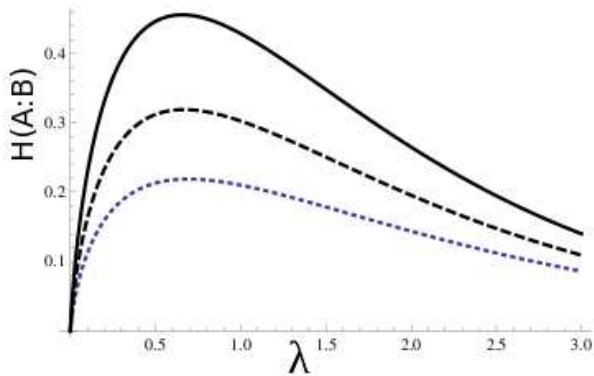}}
\caption{A plot of $H(A:B)$ as a function of $\lambda$, for different values of the efficiency.  In all of the plots $q=3.9\times 10^{-8}$, which corresponds to a dark count rate of 300/s and a time bin width of 130ps.   The solid black line is for $\eta=0.8$, the dashed line corresponds to $\eta=0.7$ while the dotted line corresponds to $\eta=0.6$.  The mutual information, $H(A:B)$, is measured units of bits.}.
\label{h1}
\end{figure}

The distribution (\ref{poisson}) in equation (\ref{twomoment2}) yields the following expression for the moment generating function
\begin{equation}
M_{loss}(\mu,\xi)=\exp\left[-\eta\lambda (\mu+\xi-\eta\mu\xi)\right].
\end{equation}
From equations (\ref{piprobs}) and (\ref{fullprobs}) we find that the joint probability distrubtion for the detectors is
\begin{eqnarray}
\mathcal{P}^{AB}(0,0)&=&(1-q)^2e^{-\lambda\eta(2-\eta)},\nonumber\\
\mathcal{P}^{AB}(c,0)&=&(1-q)e^{-\lambda\eta}-(1-q)^2e^{-\lambda\eta(2-\eta)},\nonumber\\
\mathcal{P}^{AB}(0,c)&=&=\mathcal{P}^{AB}(c,0),\\
\mathcal{P}^{AB}(c,c)&=&1-2(1-q)e^{-\lambda\eta}+(1-q)^2e^{-\lambda\eta(2-\eta)}.\nonumber
\end{eqnarray}
The marginal probabilities for Alice's (or Bob's) detector to click is thus $\mathcal{P}(0)=(1-q)e^{-\lambda\eta}$  and $\mathcal{P}(c)=1-(1-q)e^{-\lambda\eta}$.  The mutual information can easily be calculated and is found to be
\begin{eqnarray}
\label{mutin}
H(A:B)&=&2H_2\left(A\right)
+B\log B\nonumber\\
&+&2(A-B)\log(A-B)\\
&+&[1-2A+B]\log[1-2A+B],\nonumber
\end{eqnarray}
where $H_2(x)=-x\log(x)-(1-x)\log(1-x)$, $A=(1-q)e^{-\lambda\eta}$ and $B=A^2e^{\lambda\eta^2}$.  The above expression gives the mutual information between Alice and Bob as a function of the mean number of photons $\lambda$, the efficiency $\eta$ and the probability of getting a dark count $q$.  In figure \ref{h1} the mutual information is plotted as a function of $\lambda$ for different values of $\eta$, with $q=3.9\times 10^{-8}$, which corresponds to time bins of width 130ps and the detectors having on average 300 dark counts per second.  In the ideal case, $H(A:B)$ would be one bit.  From figure \ref{h1} we see that inefficiencies can significantly decrease the mutual information.  It is thus important to maximise the possible mutual information by controlling the value of $\lambda$.  The optimal value for $\lambda$ can be found using equation (\ref{mutin}).

In order to compare different experimental QKD schemes it is often useful to determine the number of bits per photon.  Equation (\ref{igen}) gives the number of bits per generated photon, while equation (\ref{idet}) gives the number of bits per detected photon.  The mutual information per generated photon, $I_g(A;B)$, is plotted in figure 2 as a function of $\lambda$.  Figure 2(a) shows that for an efficiency of 0.85, we can have greater than 10 bits per generated photon.  Figure 2(b) shows that with a lower dark count and efficiency of 0.8, one can achieve more than 13 bits per generated photon.  The information per detected photon, $I_d(A;B)$, is plotted in figure 3 as a function of $\lambda$.  One can see that in figure 3(a) we have more than 14 bits per detected photon, for $\eta=0.8$.  In figure 4(b) we find that for $\eta=0.8$ and a dark count that is about 10\% of $\lambda$, we obtain about 20 bits per detected photon.
\begin{figure}
\center{\includegraphics[width=8cm,height=!]
{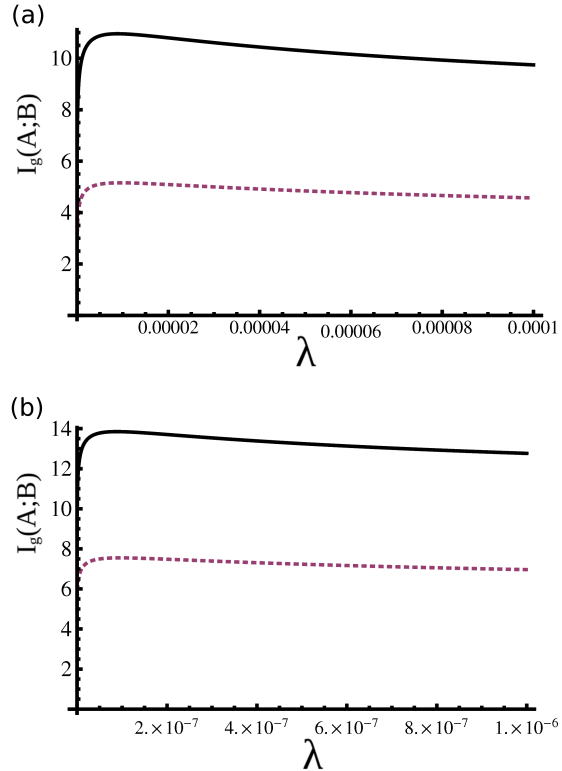}}
\caption{A plot of $I_g(A;B)$ as a function of $\lambda$. Plots (a) have $q=3.9\times 10^{-6}$ and the dotted line is for $\eta=0.6$, while the solid line is for $\eta=0.85$.  Plots (b) have $q=3.9\times 10^{-8}$ and the dotted line is for $\eta=0.6$, while the solid line is for $\eta=0.8$.  In all plots the mutual information per generated photon is measured units of bits.}
\label{fig3}
\end{figure}

\begin{figure}
\center{\includegraphics[width=8cm,height=!]
{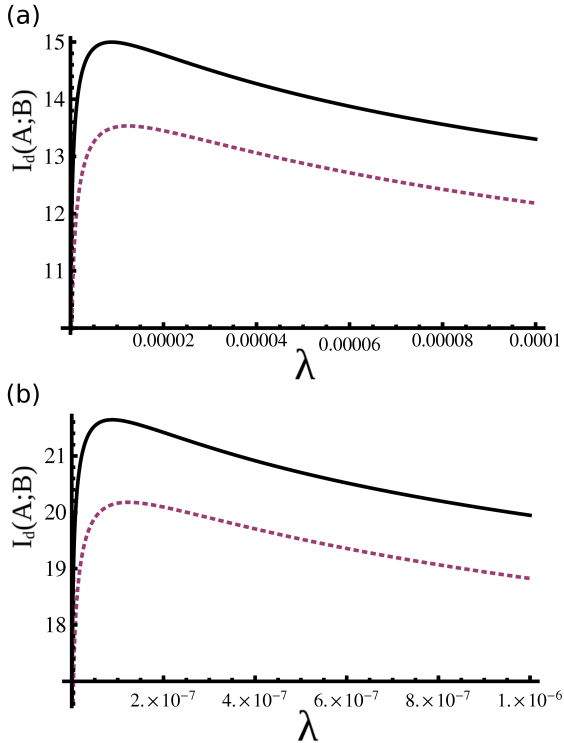}}
\caption{A plot of $I_d(A;B)$ as a function of $\lambda$. Plots (a) are for $q=3.9\times 10^{-6}$, while in plots (b) are for $q=3.9\times 10^{-8}$.  In both plots the dotted line is for $\eta=0.4$ and the solid line corresponds to $\eta=0.8$.The mutual information per detected photon is measured units of bits.}
\label{fig4}
\end{figure}

The result (\ref{mutin}) can be used with equations (\ref{igen}) and (\ref{idet}) to find the value of $\lambda$ that maximizes the information per photon.  From the perspective of experimentally implementing high bit rate QKD, one might wish to determine the laser intensity that optimizes the number of bits per photon.  This task can be achieved using our results together with the fact that the value of $\lambda$ will depend on the laser intensity \cite{ling}.  This is one of the key finding of this paper.

\begin{figure}
\center{\includegraphics[width=8.5cm,height=!]
{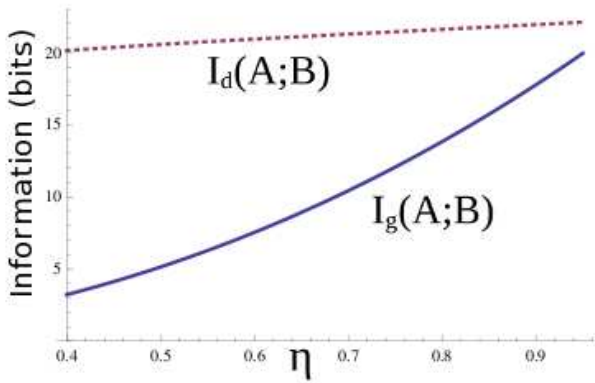}}
\caption{A plot of $I(A;B)$ as a function of $\eta$.  The probability of dark counts, $q$, is $3.9\times 10^{-8}$.  The mutual information per photon is measured units of bits.}
\label{fig7}
\end{figure}

The dependence of $I_g(A:B)$ and $I_d(A:B)$ on the efficiency can be of practical significance for implementing QKD.  For example, in an experiment one might what to know how big an improvement there would be if better detectors are used.  This problem can again be solved using equation (\ref{mutin}).  Mathematically the problem is to find the maximum of $I_g(A:B)$ or $I_d(A:B)$ for fixed values of $\eta$ and $q$.  This quantity is plotted in figure \ref{fig7}.  The plot is for a fixed value of $q$, however, decreasing $q$ does not significantly increase the information.  Similarly, the information is not decreased by too much if $q$ is increased by an order of magnitude.  If $q$ is increased by several orders of magnitude, then $I_{g,d}(A:B)$ will be decreased by an noticeable amount.  One thus sees that for reasonable small dark counts, the efficiency is the main factor limiting the information per photon.

\section{Further applications}
\subsection{Information based classification of high dimensional entangled states}
The theory we have outlined has, thus far, be applied to quantum communication.  There are, however, other situations where these results can be applied.  An example of this is in developing experimentally useful classifications of entangled states.   One approach is to use the information that two parties can extract from an entangled state as a figure of how entangled the state is \cite{indexcorr,hallcorr,hall}.  Recently, this approach has been used as the basis for an experimental procedure for quantifying the entanglement in photons generated by SPDC \cite{rochester}.  The idea was to use the fact that a pair of photons generated by SPDC are entangled both in position and momentum.  The results of measurements on each photons position or momentum would thus be correlated\footnote{To simplify the discussion we do not differentiate between correlations and anti-correlations.  The term correlated will thus encompass both situations.}.  The mutual information gained by measuring these quantities will thus give us an indication of the strength of the entanglement.

If one is to make accurate comparisons between experimental results and theory, it is important to factor in effects such as detector inefficiencies, dark counts and imperfections in the source.  This can be achieved using the formalism that we have developed.  In particular, one can see how the experimental imperfections should affect the measured entanglement of the state. 

In the experiments discussed in \cite{rochester} the measurements of position and momentum were made using spatial resolving photon detectors.  These consisted of a detector with discrete pixels.  Detection of photons by one of the pixels would thus allow for a measurement of the spatial location of the photons.  Each pixel has a finite width.  The measurement thus has a discrete set of outcomes.  One subtle point is that if the region of space is chosen sufficiently large, then the probability for detecting photons at various locations can vary.  In our analysis we implicitly assume that each outcome is equiprobable.  This assumption means that applying our theory to this experiment will lead to a slight overestimate of the mutual information.  This suggests that the proper way of viewing our results, when applied to this setup, is as providing an upper bound on the possible mutual information.

\subsection{The capacity of fibre arrays}
Another interesting application of our theory is in characterizing the information capacity of a fibre array.  This would entail treating the array as a information channel and calculating the maximum mutual information between the outputs and input with a fixed probe beam.  The capacity gives an information theoretic measure of the ability of a given array to transmit and sort optical pulses.  An example of the sort of system where this could be important is in time multiplexing of detectors \cite{timemulti,PadgettBuller}.

The two situations we consider are $M$ input fibres coupled to a single detector and a single input fibre coupled to $M$ fibres.  In the former situation we have $M$ inputs and one detector, while in the later we have one input and $M$ detectors.  One crucial difference between both of these cases and all our previous examples is that we are not considering pairs of entangled photons.  Instead, our inputs will be pulses that contain $n$ photons with probability $P(n)$.  In many instances the statistics of the input pulses can be approximated by a Poissonian distribution.  For example, if one takes each pulse to be in a coherent state \cite{loudon}.   When this is the case, the capacity can be calculated using equation (\ref{mutin}).  

\begin{figure}
\center{\includegraphics[width=8.4cm,height=!]
{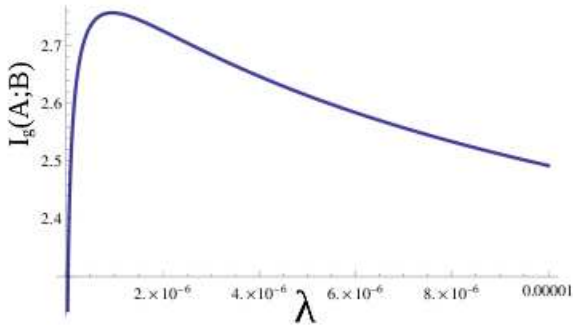}}
\caption{A plot of $I_g(A;B)$ as a function of $\lambda$.  The probability of dark counts, $q$, is $3\times 10^{-7}$ and the efficiency is $\eta=0.4$.  The quantity $I_g(A;b)$ is measured units of bits.}
\label{figarr}
\end{figure}

The approach is best illustrated by a simple example.  Suppose we have a fibre array composed of 8 output fibres coupled to a single detector.  This situation has be experimentally realized in \cite{PadgettBuller}.  The array is designed so that each fibre has a different length so that the different inputs reach the detector at different well defined times.  Let us assume that the separation between these time bins is 1ns, which is larger than the jitter of our detector.  The efficiency of the detector and array is taken to be 40$\%$, while the dark count rate is 300/s.  The probability of obtaining a dark count in a given time window is thus $q=3\times 10^{-7}$.  Finally, we assume that the input pulses have a Poissonian photon distribution with mean $\lambda$ and that the pulses are equally likely to enter each input.  Under these conditions equation (\ref{mutin}) can be used to calculate the total information as a function of $\lambda$.  In figure \ref{figarr} we plot the average information per generated photon.

The previous calculations were for a beam that has a Poissonian photon distribution.  If the beam is not Poissonian or cannot be approximated by a Poissonian, then equation (\ref{mutin}) cannot be used.  Instead, one can use the general formalism outlined in section 3.  The procedure would thus be to calculate the generating function for the given choice of $P(n)$.  Equations (\ref{piprobs}) and (\ref{fullprobs}) can then be used to calculate the Alice and Bob's joint probability distribution.  

\section{Conclusions}
We have investigated how realistic experimental conditions affect the amount of shared information that two parties can extract from entangled photons.  
A key goal of our work is to investigate systems where multiple bits can be encoded on each photon.  This means that our analysis goes beyond previous work, which has focused on encoding information in polarization.  Our approach was to construct simple but realistic models of the entangled photon source, the information channel and the detectors.  These models allowed us to take account of effects such as transmission losses, cross talk, detector inefficiencies and dark counts.  After developing the general theory in sections 3 and 4, the formalism was illustrated by looking at systems where the photon pairs are generated by spontaneous parametric down conversion.  An explicit expression for the mutual information, equation (\ref{mutin}), was given for this case.  This represented one of the main results of the paper.  Within a QKD scheme the quantity we have calculated corresponds to the shared information in Alice and Bob's keys, before privacy amplification.  Our results can thus  be used in the design of QKD experiments to choose parameters that maximise both the mutual information and the average information per photon.  As an example, figure \ref{fig7} shows the optimal amount of information per photon that two parties can share as a function of the efficiency.


Our findings have applications out with quantum key distribution.  This was demonstrated by two examples.  The first was using the mutual information as a basis for an experimental protocol to quantify photonic entanglement \cite{rochester}.  The second application was in characterizing the efficiency of an optical array in terms of how well it transmits information.  This provides a useful, experimentally accessible, figure of merit for how well an optical array can sort and transmit optical signals.

\section*{Acknowledgements}
We would like to thank Daniel Gauthier, Paul Kwiat and Kevin McCusker for very useful discussions.  This research was supported by the DARPA InPho program through the US Army Research Office award W911NF-10-0395.  SMB also acknowledges the Royal society and the Wolfson Foundation for support.

\end{document}